\documentclass[a4paper,11pt]{article}

\usepackage{contribution}

\newcommand{\weblink}[2][]{%
    \ifthenelse{\equal{#1}{}}%
    {\textnormal{\url{#2}}}%
    {\textnormal{\href{#2}{#1}}}%
}

\def\beq{\begin{equation}}
\def\eeq#1{\label{#1}\end{equation}}
\def\eeqn{\end{equation}}

\def\beqa{\begin{eqnarray}}
\def\eeqa#1{\label{#1}\end{eqnarray}}
\def\eeqan{\end{eqnarray}}

\let\bar=\overbar

\def\O{{\cal O}}

\def\Dslash{\not{\hbox{\kern-4pt $D$}}}
\def\dslash{\not{\hbox{\kern-2pt $\del$}}}

\def\msb{{\bar{\ssstyle M \kern -1pt S}}}

\newcommand{\contribution}[7][]{%
  \clearpage
  \thispagestyle{plain}
  \ifthenelse{\equal{#1}{}}
  {\hypersetup{pdftitle={#2}}}
  {\hypersetup{pdftitle={#1}}}
  \hypersetup{pdfauthor={{#3} {#4}}}
  {\centering\normalfont\LARGE\bfseries\sffamily #2 \par\nobreak}
  \lhead{}
  \chead{%
    \textit{\footnotesize XIV International Conference on Hadron Spectroscopy
      (\weblink[\textit{hadron2011}]{http://www.hadron2011.de}), 13-17 June 2011, Munich, Germany}%
  }
  \rhead{}
  \bigskip
  \begin{center}
    {#3} {#4}\ifthenelse{\equal{#6}{}}{}{\footnote{\weblink[#6]{mailto:#6}}}
    \ifthenelse{\equal{#7}{}}{}{#7} \\
    \textit{#5}
  \end{center}
  \bigskip
}

\renewcommand{\abstract}[1]{%
  \begin{center}
    \begin{minipage}{0.85\textwidth}
      \begin{footnotesize}
        #1
      \end{footnotesize}
    \end{minipage}
  \end{center}
  \bigskip
}

\begin{document}

%
%
%
%
%
{  

\makeatletter
\@ifundefined{c@affiliation}%
{\newcounter{affiliation}}%
{}%
\makeatother
\newcommand{\affiliation}[2][]{\setcounter{affiliation}{#2}%
  \ensuremath{{^{\alph{affiliation}}}\text{#1}}}
%

%

%
\contribution[Glueballs at LHC]
{Glueballs from gluon jets at the LHC}
{Wolfgang}{Ochs}  
{\affiliation[Max-Planck-Institut für Physik, 
  D-80805 Munich, GERMANY]{1} \\
 \affiliation[University of Bern,  CH - 3012 Bern, SWITZERLAND]{2}}
{}
{\!\!$^,\affiliation{1}$, Peter Minkowski\affiliation{2}}
%

\abstract{%
The existence of glueballs within QCD is uncontroversial but their
experimental verification is still in doubt. We discuss the new
possibilities for a search of glueballs as the leading object in gluon jets
at the LHC. We summarize previous results from LEP which demonstrate a
significant excess rate
of electrically neutral leading clusters in comparison with MC models.
}
%


\section{QCD expectations and search for glueballs}

According to an early prediction of QCD the self-interacting gluons 
are able to bind themselves and to give rise to a new 
spectroscopy of gluonic matter or glueballs; specific scenarios for glueball
phenomenology go back to 1975 \cite{Fritzsch:1975tx}, for 
recent reviews, see \cite{Klempt:2007cp}. 
Today, quantitative 
predictions are derived from lattice QCD and from QCD sum rules. The
lightest glueball is found to be a scalar with $J^{PC}=0^{++}$ 
and a mass of 1.0-1.7 GeV.

There have been considerable efforts to identify glueballs experimentally.
 The aim is at first to establish the
lightest $q\bar q$ nonets in the spectrum; then, the appearence of
extra states could hint towards a glueball. More directly, one looks for
an enhanced production of a glueball candidate 
in "gluon rich" processes but a suppression in $\gamma\gamma$ reactions
 \cite{Klempt:2007cp}.

Studies of production and decay of resonances along these lines have
led to various scenarios for classification of the lightest scalar states
$$f_0(600),\ f_0(980), \ f_0(1370),\ f_0(1500),\ f_0(1710).$$
According to one approach (e.g. \cite{Amsler:1995tu}), the glueball
could be mixed into the three states above 1 GeV together with two members
of the $q\bar q$ nonet. Alternatively, the glueball could be related to the
broad $f_0(600)$ (e.g. \cite{Minkowski:1998mf,narison}).
A problem in constructing multiplets is the status of 
$f_0(1370)$ which is
not seen in phase shift analyses of 
$\pi^+\pi^-\to \pi^+\pi^-, \pi^0\pi^0$ \cite{Ochs:2010cv}.

The production of resonances has been studied
in a number of gluon-rich processes:
"central production" $pp\to p\ gb\ p$ by double
Pomeron exchange,  $J/\psi \to \gamma\ gb$,
$p\bar p \to \pi\ gb$, $B\to K\ gb$ (through $b\to s g$) and, finally,
 forward fragmentation of a gluon into glueballs. 


Only in the last reaction involving a high energy gluon jet the gluon 
can be identified 
as a source, in the other processes the overall energy is low of \O\  (few GeV) 
and the role of gluons is not obvious anymore. An interesting result on central
production has been presented at this conference by ALICE at LHC 
\cite{schicker}: in the double gap events the isoscalar states
$f_0(980)$ and $f_2(1270)$ are enhanced in comparison to no-gap events.
The enhanced production of the well known $q\bar q$ state $f_2(1270)$ 
demonstrates that the double Pomeron mechanism does not enhance exclusively 
glueballs. We also note that the Pomeron structure has been
investigated at HERA \cite{Chekanov:2005vv}. 
 Present results suggest a dominant fracton ($\sim
70\%$) of
momentum to be carried by gluons, but the ratio
$G(x)/S(x)$ of gluon
over singlet quark densities at large momentum fractions $x\sim 1$ 
varies
between $G/S\sim 0$ for ZEUS data according to
\cite{Royon:2006by}
 and  $G/S\sim 1-2$ for H1. 
Then, the production of $q\bar q$ states in double Pomeron scattering could
be a reflection of the sizable quark valence component of the Pomeron.

\section{Leading particle systems in gluon jets}
According to the well known concept of quark fragmentation 
the leading particles
at large Feynman~$x$ are those which carry the initial quark of the jet as a
valence quark
$$ q\to \text{Meson}\ (q\bar q')+X, $$
for example, leading particles in a $u$-quark jet are a $\pi^+(u\bar d)$ 
or a $\pi^0(\{u\bar u +d\bar d\}/\sqrt{2})$ 
with half strength, whereas  $\pi^-(d\bar u)$ is
suppressed at large~$x$. In analogy, one can consider the 
fragmentation of a gluon 
and suppose that the leading particle in the jet is the one with a gluonic
valence component
$$g \to \text{Meson}\ (gg) +X.$$
Models of this kind with leading glueballs, but also with leading
isoscalars like $\eta,\ \eta',\ \omega$ at large $x$
have been suggested already long ago
\cite{Peterson:1980ax}, for $x$-distributions,
see also \cite{Roy:1999bu}.

Studies of gluon jets at LEP did not establish a clear support of the
model for isoscalar $q\bar q$ mesons \cite{Acciarri:1995yp}.
While the L3 collaboration found for the jet of lowest momentum in
$e^+e^-\to $ 3 jets a considerable
enhancement of $\eta$ production by factor 2-3 beyond MC calculations 
 for $x_p>0.06$, where $x_p=p/p_{beam}$, ALEPH later reported agreement
between data and revised MC versions (similarly also for $\eta'$, but
with low statistics). OPAL found an
excess $\eta$ rate over MC's at the higher momenta but did
not separate quark and gluon jets in this range. No other isoscalar 
particles have been studied separately for quark and gluon jets.

The distributions of charge and mass of the leading cluster  $Q_{lead}$ and
$M_{lead}$ in  
gluon jets beyond a rapidity gap reflect the colour 
neutralisation mechanism \cite{Minkowski:2000qp}. In particular,
the ``color octet neutralisation'' is a precondition for
glueball production. In that case, two gluons, if
separated beyond the confinement radius $R_c$, will create two other gluons
to form colour neutral sub-systems. Alternatively, colour triplet neutralisation
is possible with creation of two $q\bar q$ pairs, or both mechanisms with
probablilities $P_8$ and $P_3$. For large rapidity gaps one expects the
charge distribution to approach a limiting behaviour with charge
$Q_{lead}=0$ with $P_8$ and charges $Q_{lead}=0,\pm1$ with $P_3$,
 
Results from LEP on leading clusters have been obtained from OPAL,
DELPHI 
and ALEPH
\cite{Abbiendi:2003ri}.
All experiments agree upon a significantly enhanced 
rate for neutral
clusters  ($Q_{lead}=0$) beyond a rapidity gap in gluon jets by 10-40\%
as compared to the JETSET MC, depending on the selection
and purity of the jets. On the other hand, the corresponding
distributions for quark jets or for gluon jets without gap agree well with
MC's.
In addition, DELPHI and OPAL find the excess of gluon jets with $Q_{lead}=0$
with typically lower masses $M_{lead}\lesssim 2.5 $ GeV. A natural
explanation would be a leading gluonic system or glueball.

\section{Proposal for LHC studies of gluonic systems}
Studies of leading particle systems can be performed at the LHC with some advantages.
Most importantly, gluon jets can be selected with higher energies 
in comparison to LEP and they are more copiously produced with good
statistics. It is possible to compare quark and gluon jets with similar
energies from different processes.\\
\begin{table}[tb]
\begin{center} 
\begin{tabular}{|lc|cccc|}
\hline
           &       & $p_T$ & $x_T$ & g in di-jet & q in $\gamma + $ jet\\
\hline
TEVATRON  & 1.8 TeV & 50 & 0.056 & 60\% & 75 \% \\
LHC  & 7 TeV & 200 & 0.057 & 60\% & 80 \% \\
                     &       & 50  & 0.014 & 75\% & 90 \% \\
                     &       & 800 & 0.229 &  25\% &75\% \\ \hline
\end{tabular}
\caption{Rates for gluon and quark jets at TEVATRON \cite{Acosta:2004js} and LHC
\cite{Gallicchio:2011xc}. \vspace{-0.3cm}}
\label{tab:qgrates}
\end{center}
\end{table}
1. leading order processes to be calculated from $pdf's$ and parton-parton
cross sections:\\
Quark jets can be obtained from final states $pp\to \gamma+jet +X$ with 
subprocess $qg\to \gamma q$ dominating at the lower $p_T$.
Gluon jets are found among di-jet events, also at low $p_T$. Examples are
presented in Tab. 1.\\ 
A good purity of quark jets can be obtained in this way, but gluon jets with
their steeper fragmentation need higher purity to reduce background.\\
2. gluon bremsstrahlung:\\
Using a trigger on total transverse energy one selects 3-jet events. Similar
to the case at $e^+e^-\to 3\ jets$ the lowest momentum jet is most likely a
gluon jet from QCD bremsstrahlung ($qqg$, $qgg$ or $ggg$). The fraction
of gluon jets can be derived within the DGLAP approximation for low $k_T$. 
For example, at small $x_g$ one finds, given the ratio of gluon to quark production rate
$R_g=\sigma_g/\sigma_q$ the fraction 
$
F_g(x_g)=\frac{1}{1+4x_g/(8+18R_g)}
$
which yields for $R_g=1$, as example, 
$$F_g\approx 95\%\ \text{at}\ x_g=0.2, \qquad F_g\approx 85\%\ \text{at}\ x_g=0.5.$$

Studies at LHC could be useful in two directions:\\
1. Leading clusters with larger rapidity gaps.\\ 
The new possibilities at LHC follow from the gluon jet energies larger by an
order of magnitude as compareed to LEP (typically $<25$ GeV).  This allows
for a better separation of the leading cluster.  The rapidity gaps could
extend up to $\Delta y\sim 4$ (add $\ln 10\approx 2.3$ to $\Delta y\approx
1.7$).
With increasing 
rapidity gaps the leading charges should be closer to their
asymptotic distribution with values $Q=0,\pm 1$ allowing for a 
better estimate of probabilities $P_3$ and $P_8$. \\
2. Direct study of resonances in mass distributions.\\
The spectra of the invariant mass of leading particles beyond the gap 
are generally found quite smooth. 
There is some evidence in gluon jets for $f_0(980)$ in the $\pi\pi$
(DELPHI 
\cite{Abbiendi:2003ri}) 
and for $f_0(1500)$ in the $4\pi$ spectrum
(OPAL \cite{Abbiendi:2003ri}).  The rapidity gap cuts affect the angular decay
distribution of the cluster and could reduce the resonance signal.  This is
avoided if the mass spectra are constructed first and then their $x$
distribution is studied. Such resonance $x$ spectra have not yet been determined
for quark and gluon jets separately.  

The distinguishing feature for identifying a glueball is its $x$
distribution in comparison with the reference 
spectrum of a well defined $q\bar q$
resonance (examples $ \rho,\ f_2(1270), \ \phi(1020)$) of comparable mass
in the quark and gluon jet resp. The glueball should be "suppressed" in a
quark jet and should be "leading" in a gluon jet, i.e. above a $q\bar q$
reference resonance, according to the following scheme (for scalar $f_0$'s): 

\begin{table}[h]
\begin{center}
\begin{tabular}{|l|lll|}
\hline
meson          & quark jet & gluon jet      &  \\
               &           & triplet neutr. & octet neutr.\\
\hline
{ $q\bar q: f_0\ [ref:\rho,f_2,\phi\ldots]$} & \underline{ leading} & suppressed &
suppressed\\
{ $gb: f_0\  [ref:\rho,f_2\ldots]$}         & suppressed&suppressed&\underline{leading}  \\
{ $q\bar q: f_0$, strongly mixed}& \underline{ leading}& suppressed& leading
(?)\\
{$4q: f_0(600)/\sigma, f_0(980)$ (?)}  & suppressed & suppressed & suppressed\\
\hline
\end{tabular}
\caption{Identifying glueballs through reference spectra in quark and gluon
jets.\vspace{-0.3cm}}
\end{center}
\end{table}
The last line in the table relies on the validity of particular models,
here we refer to the quark counting approach \cite{Brodsky:1973kr}. 
To the extent, that the $x$-distribution of $f_0(980)$ almost coincides with
the one of $f_2(1270)$ \cite{Ackerstaff:1998ue}, there is no evidence for structure beyond $q\bar q$
of $f_0(980)$.

In principle, also mixed $gb-q\bar q$ states could be recognized by
comparing with the appropriate superposition of two reference distributions.
Gluonic components could appear in the spectra of $(\pi\pi)^0$
($f_0(600)/"\sigma",\ f_0(980),\ f_0(1500)$), of $(4\pi)^0$ 
($(f_0(1370)(?),\ f_0(1500)$) and $(K\bar K)^0$, ($ f_0(980),\ f_0(1500), \
f_0(1710)$). 

\section{Summary}
1. The existence of glueballs is predicted since long, the clear evidence
is still missing.

2. Lesson from LEP: evidence for a new fragmentation component:\\ 
excess of neutral clusters by up to 40\% beyond expectation from MC's.
\\
Gluon jets may not be built up from quark strings only.  

3. There is a new chance of finding glueballs in gluon jets at LHC:\\ 
excess of neutral leading clusters with increasing gap size;\\ 
resonance $x$-spectra  in quark and gluon jets in comparison with reference
spectra.

%

}  



\end{document}